\documentclass{aa}
\usepackage{graphicx}
\usepackage{txfonts}
\begin{document}

\titlerunning{Rotating CAK topological analysis}
\authorrunning{M. Cur\'{e} \& D. Rial}
\title{The influence of rotation in radiation driven winds from hot stars}
\subtitle{II. CAK topological analysis} 
\author{Michel Cur\'{e} \inst{1}
\and Diego F. Rial \inst{2}}
\offprints{michel.cure@uv.cl}
\institute{Departamento de F\'{\i}sica y Meteorolog\'{\i}a, Facultad de Ciencias,  
Universidad de Valpara\'{\i}so, Valpara\'{\i}so, Chile
\and Departamento de Matem\'{a}ticas, Facultad de Ciencias Exactas y Naturales, 
Universidad de Buenos Aires, Argentina}  
\abstract{
The topological analysis from Bjorkman~(1995) for the standard model that describes the winds from hot stars by Castor, Abbott \& Klein~(1975) has been extended to include the effect of stellar rotation and changes in the ionization of the wind.
The differential equation for the momentum of the wind is non--linear and transcendental for the velocity gradient. Due to this non--linearity the number of solutions that this equation possess is not known. After a change of variables and the introduction of a new physically meaningless independent variable, we manage to replace the non--linear momentum differential equation by a system of differential equations where all the derivatives are {\it{explicitely}} given. We then use this system of equations to study the topology of the rotating--CAK model. For the particular case when the wind is frozen in ionization ($\delta=0$) only one physical solution is found, the standard CAK solution, with a X--type singular point. For the more general case ($\delta \neq 0$), besides the standard CAK singular point, we find a second singular point which is focal--type (or attractor). We find also, that the wind does not adopt the maximal mass--loss rate but almost the minimal.
} 
\maketitle
\section{Introduction} 

Since the launch of the first satellite with a telescope on board, it has been established the widespread presence of stellar winds from hot stars. These winds are driven by the transfer of momentum of the radiation field to the gas by scattering of radiation in spectral lines (Lucy and Solomon, 1970). The theory of radiation driven stellar winds is the standard tool to describe the observed properties of the winds from these stars. Castor, Abbott and Klein (1975, hereafter "CAK") obtained an analytical hydrodynamic model for these winds, based in the Sobolev approximation.
The CAK model has been improved by Friend and Abbott (1986,"FA") and Pauldrach et al. (1986,"PPK"), giving a general agreement with the observations. For a extended review see Kudritzki and Puls (2000, "KP") and references therein.

This agreement with the observations led to the development of a new method to
determine galactic distances using Supergiants as targets, namely the Wind Momentum Luminosity relationship ("WML", Kudritzki et al. ~1999, KP and references therein). 

More detailed studies from Puls et al. (1996) and Lamers \& Leitherer (1993) came to the conclusion  that the line--driven wind theory shows a systematic discrepancy with the observations. Lamers \& Leitherer (1993) suggest that this discrepancy may arise due to an inadequate treatment of multiple scattering. Abbott \& Lucy (1985), Puls (1987) and Gayley et al.~(1995) have shown that multiple scattering can provide an enhancement of the wind momentum over that from single scattering only by a factor of two -- three for O stars (Abbott \& Lucy~(1985) found a factor of 3.3 for the wind of $\zeta$ Pup). 

Vink et al. (2000) calculate, including the multiple scattering effects, mass--loss rates for a grid of wind models that covers a wide range of stellar parameters. They found a much better agreement between theory and observation, concluding that the inclusion of multiple scattering increases the confidence of the WML relationship to derive extragalactic distances.
In all the calculations involved in the WML relationship, the solution of the improved (or modified) CAK wind (hereafter m--CAK) is \textit{not} used. Instead an ad--hoc $\beta$--field velocity profile is utilized (see KP, Vink et al. 2000).

The unsatisfactory results of the velocity field obtained from the m--CAK model
when applied in the WML relationship could come from the complex structure of this
non--linear transcendental equation for the velocity gradient and its solution
schema. Due to this non--linearity in the momentum differential equation, there exist many 
solution branches in the integration domain. A physical solution that describes the observed 
winds must start at the stellar photosphere, satisfying certain boundary condition and reach infinity. There is no solution branch that covers the whole integration domain, thus a solution must pass through a singular point in order to match a second solution branch. Therefore, the solution in this second solution branch reaches infinity. To find the location of singular points is one of the most difficult aspects of topological analysis of non-linear differential equations.

Bjorkman~(1995) performed a topological analysis of the CAK differential equation. He showed that 
the original solution from CAK, which passes through a X--type critical point and has a monotonically 
increasing velocity field, is the only physical solution that satisfies the condition of 
zero pressure at infinite radius. In this study Bjorkman did \textit{not} include the influence of 
the star's rotational speed.

Although it is known that the line--force parameters (see below) are not constant through in the wind (Abbott 1982), the standard m--CAK model still uses these parameters as constant. For the particular  case of extreme low metalicities, Kudritzki (2002) introduced a new treatment of the line--force with depth dependent radiative force multipliers. As a test, he applied this new treatment for the most massive and most luminous O stars in the Galaxy and in the Magellanic Clouds (due to the lower metalicity) finding an acceptable agreement between theory and observations. Then, it was used to understand the influence of stellar winds on the evolution of very massive stars in the early universe and on the interstellar medium in the early stages of galaxy formation.

On the other hand, it is known from observations, that all early type stars have moderate to large rotational speeds (Hutchings et al. 1979, Abt et al. 2002) and for Oe and Be stars, their rotational speed is a large fraction of their break--up speed (Slettebak 1976,  Chauville et al. 2001). 
The incorporation of rotation in the CAK and m--CAK models has been studied by Castor (1979), 
Marlborough \& Zamir (1984), FA and PPK, concluding that the effect of the centrifugal force 
results in a downstream--shift of the position of the singular point, a slightly lower terminal 
velocity and a slightly larger mass loss rate as a result of an increasing in the star's 
rotational speed. 
Maeder (2001) studied the influence of the stellar rotation in the WML relationship, finding just 
a very small effect on it.

A revision of the influence of the stellar rotation in radiation driven winds has been done by 
Cur\'{e} (2004) finding that there exists a second singular point in these winds. He studied the 
case when the stellar rotational speed is high and found numerical solutions, that pass through 
this second singular point, and which are denser and slower than the standard m--CAK solution.

In view of these results, it is crucial to understand the {\it{solution topology}} of the standard 
model, forall when one wants to incorporate other physical processes into the theory.

The purpose of this article is to study the topology of the rotating--CAK model. In section 
\ref{sec-HYD} we give a brief exposition of the radiation driven winds theory and 
the non--linear differential equation for the momentum, including rotation, is shown. 
In section \ref{sec3} after a coordinate transformation, we develop a general method 
that allows to replace the non--linear momentum equation in a simple and 
straightforward manner by a system of ordinary differential equations, where all 
the derivatives  are explicitely given. In section \ref{sec4} a general condition for 
the eigenvalue of the problem is developed. This condition allows to classify the 
topology of the singular point (Saddle or Focal) and constrains the location of it 
in the integration domain. Section \ref{sec5} is devoted to the application of the 
criteria developed in section \ref{sec4} for the rotating--CAK model. In section 
\ref{sec6} we show numerical results of this topological analysis, first for 
a wind frozen in ionization (setting the $\delta$ parameter of the line--force to 
zero) and compare our results with Bjorkman`s (1995) for the non--rotating CAK model
and the rotating--CAK model from Marlborough \& Zamir (1984). Furthermore, section 
\ref{sec6} shows the results of the influence of changes in the wind`s ionization 
structure ($\delta \neq 0$) on the topology and discuss the rotating--CAK wind model.
Conclusion are summarized in section \ref{sec-CONC}.

\section{The non--linear differential equation \label{sec-HYD}} 
The standard stationary model for radiation driven stellar winds treats an one--component 
isothermal radial flow, ignoring the influence of heat conduction, viscosity and magnetic 
fields (see e.g., Kudritzki et al. 1989, "KPPA").

For a star with mass $M$, radius $R_{\ast}$, effective temperature $T$ and
luminosity $L$, the momentum equation with the inclusion of the centrifugal force due to star's 
rotation, reads:

\begin{equation} 
v\frac{dv}{dr}=-\frac{1}{\rho }\frac{{d\wp}}{dr}-\frac{G M (1-\Gamma )}{r^{2}}+ 
\frac{v_{\phi }^{2}(r)}{r}+g^{line}(\rho, v', n_{E})  \label{2.1} 
\end{equation} where 
$v$ is the fluid velocity, $v' = dv/dr$ is the velocity gradient, $\rho$ is the 
mass density, $\dot{M}$ is the star's mass loss rate, $\wp$ is the fluid pressure, 
$v_{\phi } = v_{rot} \, R_{\ast} / r$, where $v_{rot}$ is the star's
rotational speed at the equator, $\Gamma$ represents the ratio of the radiative
acceleration due to the continuous flux mean opacity, $\sigma_{e}$, relative to
the gravitational acceleration, i.e., $\Gamma= \sigma_{e} L / 4 \pi c G M$ and
the last term $g^{line}$ represents the acceleration due to the lines. \\

The standard parameterization of the line--force (Abbott, 1982) reads: 
\begin{equation} 
g^{line}=\frac{C}{r^{2}}\;f_{D}(r,v,v')\;\left( r^{2}v\; v'\right) ^{\alpha
}\;\left( {n_{E}}/{W(r)}\right) ^{\delta }  \label{2.2} 
\end{equation}
$W(r)$ is the dilution factor and $f_{D}(r,v,v')$ is the finite disk correction factor. The 
line force parameters are: $\alpha$, $\delta$ and $k$ (the last has been incorporated in the 
constant $C$), typical values of these parameter from LTE and non--LTE calculations are 
summarized in Lamers \& Cassinelli (1999, chapter 8, "LC"). 
We have not used the absolute value of the velocity gradient in the line--force 
term, because we are interested in monotonic velocity laws.\\
The constant $C$ represents the 
eigenvalue of the problem (see below) and is given by:
\begin{equation} 
C=\Gamma G M k \left( \frac{4\pi }{\sigma _{E} v_{th} \dot{M}}\right) 
^{\alpha }\;\label{2.3},
\end{equation} where $v_{th}=({2k_{bol}T/m_{H}})^{1/2}$ is the hydrogen thermal speed, $n_{E}$ 
is the electron number density in units of $10^{-11}cm^{-3}$ (Abbott 1982), while the meaning 
of all other quantities is standard (see, e.g., LC). 

Together with the momentum equation (\ref{2.1}), the continuity equation reads:
\begin{equation} \;
4 \pi r^{2}\rho \;v = \dot{M}  \label{2.0} 
\end{equation}

Introducing the following change of variables:
\begin{equation} 
u=-R_{\ast}/r  \label{2.4a} ,\\
\end{equation}
\begin{equation} 
w= \; v/a  \label{2.4b} ,\\
\end{equation}
\begin{equation} 
w'=\; dw/du  ,\label{2.4c}
\end{equation} where $a$ is the isothermal sound speed, i.e., $\wp=a^{2}\rho$. Replacing the density $\rho$ from (\ref{2.0}), the m--CAK momentum equation (\ref{2.1}) with the line force (\ref{2.2}), we obtain: 
\begin{eqnarray} 
F(u,w,w^\prime)&\equiv &\left( 1-\frac{1}{w^{2}} \right) w \; w'+ \frac{2}{u} + A \; \left( 1+\Omega^{2}\,u \right) \nonumber \\
               & & -\; \overline{C} \;f_{D}\;g(u)(w)^{-\delta }\left( w \;w'\right)^{\alpha}\ = \; 0 \label{2.5} 
\end{eqnarray} 
here the constants $A$, $\overline{C}$ and $\Omega$ are:
\begin{equation} 
A = \frac{GM(1-\Gamma )}{a^{2}R_{\ast }}=\frac{v_{esc}^{2}}{2a^{2}} =\frac{v_{break-up}^{2}}{a^{2}} 
 \label{2.5d}, 
\end{equation} 
\begin{equation} 
\overline{C} = C\;\left( \frac{\dot{M} D}{2\pi}\frac{10^{-11}}{aR_{\ast
}^{2}} \right)^{\delta }\;(a^{2}R_{\ast })^{(\alpha -1)}  \label{2.5b} ,
\end{equation} 
\begin{equation} 
\Omega=\frac{v_{rot}}{v_{break-up}}= \frac{a_{rot}}{\sqrt{A}} , \label{2.5f} 
\end{equation} where $a_{rot}$ is defined by: 
\begin{equation} 
a_{rot}=\frac{v_{rot}}{a} , \label{2.5e} 
\end{equation}
here $v_{esc}$ is the escape velocity and $v_{break-up}$ is the "break--up" velocity. The function $g(u)$ is defined as:
\begin{equation} 
g(u)=\left( {1+\sqrt{1-u^{2}}}\right)^{\delta } ,  \label{2.5c} 
\end{equation} and the constant $D$ is:
\begin{equation} 
D= \frac{1}{m_{H}} \frac{(1+Z_{He} Y_{He})}{(1+A_{He} Y_{He})} \label{2.5g} ,
\end{equation} $Y_{He}$ is the helium abundance relative to the hydrogen,
$Z_{He}$ is the amount of free electrons provided by helium, $A_{He}$ is the atomic mass number of helium and $m_{H}$ is the mass of the proton.

The standard solution, from this non-linear differential equation
(\ref{2.5}), starts at the stellar surface and after crossing the
singular point reaches infinity. At the stellar surface the differential
equation must satisfy a boundary condition, namely the monochromatic
optical depth integral (see Kudritzki, 2002, eq. [48]):
\begin{equation} 
\tau_{Phot}= \int_{R_{\ast}}^{\infty} \sigma_{E} \,\rho \,(r) \,dr \, =\, \frac{2}{3}
  \label{2.5a} 
\end{equation}
A numerically equivalent boundary condition is to set the density at the stellar surface to a specific value, 
\begin{equation} 
\rho\,(R_{\ast}) =  \rho_{\ast}. \label{2.5ab} 
\end{equation}

When the singularity condition, 
\begin{equation} 
\frac{\partial}{\partial w'}F(u,w,w')= F_{w'} = 0  \label{2.6} 
\end{equation} 
is satisfied, its location ($u=u_{c}$) corresponds to a singular (or critical) point, and in order to get a physical solution, the regularity condition must be imposed, namely:
\begin{equation} 
\frac{d}{du} F(u,w,w')=\frac{\partial F}{\partial u}+\frac{\partial 
F}{\partial w}w'= F_{u}+ F_{w}\,w'=0 , \label{2.7} 
\end{equation} hereafter, all partial derivatives are written in a shorthand form, i.e. $F_{u}={\partial F}/{\partial u}$.

In order to satisfy simultaneously equations (\ref{2.5}), (\ref{2.6}), (\ref{2.7}) and (\ref{2.5a}) or (\ref{2.5ab}), the value of the constant $\overline{C}$ is not arbitrary, i.e., the constant $\overline{C}$ is the \textit{eigenvalue} of this non--linear problem.

\section{Coordinate Transformation and System of Equations\label{sec3}}

In this section we will apply another coordinate transformation and introduce a new independent variable, $\tau$, without physical meaning. This will allow us to transform the non--linear differential equation for the momentum (\ref{2.5}) into a system of coupled differential equations, which is numerically integrable.

Defining
\begin{equation}
y = \frac{1}{2} \;w^{2} ,   \label{2.7a}
\end{equation} and
\begin{equation}
p =w\;w'=\frac{dy}{du} ,\label{2.8}
\end{equation}
the momentum equation (\ref{2.5}) can be written in a general form as:
\begin{equation}
F(u, y, p) = 0.
\end{equation}
Differentiating this function and using $dy=p\,du$, we obtain:
\begin{equation}
dF=\left(F_{u}\,+ p\,F_{y}\right) \,du + F_{p}\,dp = 0  \label{dF}
\end{equation}
We introduce now a new independent variable, $\tau$, defined implicitely by:
\begin{equation}
du=F_{p}\,d\tau  , \label{taudef}
\end{equation}
Because $u$ and $\tau$ are independent variables, they have to be related between them. 
While $F_{p}\neq 0$, we can write $\tau$ as function of $u$ as:
\begin{equation}
\frac{dy}{du}=\frac{dy}{d\tau} \left(\frac{du}{d\tau}\right)^{-1}\, = \;p
\end{equation}
We can transform from (\ref{dF}) to the following system of ordinary differential equations:
\begin{equation} 
\frac{du}{d\tau} \equiv U= F_{p} , \label{sist1} 
\end{equation}
\begin{equation}
\frac{dy}{d\tau} \equiv Y=p\,F_{p} , \label{sist2}
\end{equation} 
\begin{equation}
\frac{dp}{d\tau} \equiv P=-\left(  F_{u}\,+p\,F_{y}\right). \label{sist3}
\end{equation}
A solution of this system of differential equations is also a solution of the original momentum equation, since if any initial condition $(u_{0},y_{0},p_{0})$ satisfies $F(u_{0},y_{0},p_{0})  =0$, then a solution of (\ref{sist1}--\ref{sist3}) verifies that $F(u(\tau),y(\tau),p(\tau))=0$. \\
An advantage of this equation system (\ref{sist1}--\ref{sist3}) over the CAK momentum differential equation (\ref{2.5}) is that all the derivatives are {\it{explicitely}} given, therefore there is no need to use root--finding algorithms to find the value of the velocity--gradient. Also, standard numerical methods (e.g., Runge--Kutta) can be used to integrate this system.
\section{Linearization and Eigenvalue Criteria \label{sec4}}

All critical points of the system  (\ref{sist1}-\ref{sist3}) satisfy simultaneously $F=0$, $U=0$ and $P=0$. Thus, in order to study the behavior of the solution in the neighborhood of a singular point we linearise this system of differential equations, using the Groebman--Hartman theorem ("GH", see appendix. For more details see Amann 1990), we obtain: 
\begin{eqnarray}
dU &  =\left(  U_{u}\,+pU_{y}\right)
\,du+U_{p}\,dp,\\
dP &  =\left(  P_{u}\,+pP_{y}\right)
\,du+P_{p}\,dp.
\end{eqnarray}
We have not included $dY$ because $U$ and $Y$ are dependent ($Y=pU$, eq. \ref{sist2}).
The GH theorem indicates that the eigenvalues (and the eigenvectors) of the partial derivative matrix $B$, defined by:
\begin{equation}
B=\left(
\begin{array}
[c]{cc}
U_{u}+p\,U_{y} & U_{p}\\
P_{u}+p\,P_{y} & P_{p}%
\end{array}
\right) \, , \label{mtxB}
\end{equation} provide the information concerning the topology structure of the critical (singular) points.\\
 
On the other hand, if we consider the eigenvalue $\bar{C}$ as a free paramenter, the critical points are the 
solution of the following system of equations:
\begin{eqnarray}
F\left(u,y,p,\bar{C}\right)& = 0\,, \\
U\left(u,y,p,\bar{C}\right)& = 0\,, \\
P\left(u,y,p,\bar{C}\right)& = 0\,.
\end{eqnarray}
In this case the number of incognits is greater than the number of equations, then we can only solve for the incognits in terms of one of them (implicit function theorem). 

If $\left(u_{c},y_{c},p_{c},\bar{C}\right)$ satisfies the previous system and furthermore, $\Delta \neq 0$ 
at the singular point, we can solve for $y_{c}$, $p_{c}$ and $\bar{C}$ and its derivatives as a function of $u_{c}$. We obtain for the gradient of $\bar{C}$: 
\begin{equation}
\frac{d\bar{C}}{du_{c}}=-\frac{1}{\Delta}\det\left(
\begin{array}
[c]{ccc}%
F_{y} & F_{p} & F_{u}\\
U_{y} & U_{p} & U_{u}\\
P_{y} & P_{p} & P_{u}%
\end{array}
\right), \label{dcdu}%
\end{equation} where the determinant $\Delta$ is defined by:
\begin{equation}
\Delta=\det\left(
\begin{array}
[c]{ccc}%
F_{y} & F_{p} & F_{\bar{C}}\\
U_{y} & U_{p} & U_{\bar{C}}\\
P_{y} & P_{p} & P_{\bar{C}}
\end{array}
\right),
\end{equation}

Considering that at the critical point: $F_{p}=0$ and $F_{u}=-pF_{y}$, 
equation (\ref{dcdu}) transforms to:
\begin{equation}
\frac{d\bar{C}}{du_{c}}=-\frac{1}{\Delta}\det\left(
\begin{array}
[c]{ccc}%
F_{y} & 0 & -pF_{y}\\
U_{y} & U_{p} & U_{u}\\
P_{y} & P_{p} & P_{u}%
\end{array}
\right)  =\frac{F_{y}}{\Delta}\det\left(  B\right)  \label{dCdu2} \,.
\end{equation}
The GH theorem establishes that the singular point topology is determined by the sign of the 
eigenvalues of the matrix $B$. A critical point is of X--type (Saddle) when the eigenvalues 
are both real and have opposite sign, or equivalently,
\begin{equation}
\det(B) \, < \,0 \,. \label{dbneg}
\end{equation}
Applying this to the rotating--CAK model (see next section for details) it is verified that,
\begin{equation}
\frac{F_{y}}{\Delta}  \, >  \,0 \,\,,
\end{equation}
then a X--type singular point corresponds to the condition: 
\begin{equation}
\frac{d\bar{C}}{du_{c}} \,< \,0 \,.  \label{condcu}
\end{equation}
Therefore any X--type physically relevant singular point, is related to the behavior of the 
eigenvalue $\bar{C}$, specifically when condition (\ref{condcu}) holds.
\section{The Topology of the Rotating--CAK Model \label{sec5}}
The analysis we have shown in the last section (up to eq. \ref{dbneg}) is valid for the general radiation driven stellar winds theory, i.e., including the finite--disk correction factor.
In the remainder of this paper we will study the original CAK model, i.e. $f_{D}=1$. The topological study of the m--CAK model will be the scope of a future article. 

The non--linear momentum equation (\ref{2.5}) for the rotational--CAK model (including $\delta$), 
in ($u$, $y$, $p$) coordinates reads:
\begin{eqnarray}
F(u,y,p) & = &\left(  1-\frac{1}{2y}\right)  p+\frac{2}{u}+A\left(1+\Omega^{2}u\right) \nonumber\\
 & &-\bar{C}\,g\left(  u\right)  \,y^{-\delta/2}p^{\alpha} = 0 \,.\label{Fcak}%
\end{eqnarray}
From this equation, after a straightforward calculation, the corresponding system of differential equations $U$ and $P$, (eqs. \ref{sist1} and \ref{sist3}, respectively) are:
\begin{equation}
U\left(  u,y,p\right) = 1-\frac{1}{2\,y}-\alpha\,\bar{C}\,g(u)\,y^{-\delta/2}\,p^{-1+\alpha}\,\, \label{Ucak} ,
\end{equation}
\begin{eqnarray}
P\left(u,y,p\right) &=&\frac{2}{u^{2}}-\Omega^{2}A-\frac{p^{2}}{2\,y^{2}}\nonumber\\
& &-\frac{\delta}{2}\bar{C}\,g\left(  u\right)\,y^{-1-\delta/2}\,p^{1+\alpha}\nonumber\\
& &+\bar{C}\,g\left(  u\right)  \,h(u)\,\,y^{-\delta/2}\,\,p^{\alpha} \label{Pcak},
\end{eqnarray} where 
\begin{equation}
h\left(  u\right)  =-\,\delta\,u\left(1-u^{2}+\sqrt{1-u^{2}}\right)^{-1} \,, \label{eqh}
\end{equation}
$F_{y}$ and $\Delta$ are:
\begin{equation}
F_{y}=\frac{p}{2\,y^{2}}+\frac{\delta}{2}\bar{C}\,g\left(  u\right)
\,\,y^{-(1+\delta/2)}\,\,p^{\alpha} ,%
\end{equation} 
\begin{eqnarray} 
\Delta   & = & (1-\alpha)\,g(u)\,y^{-\delta/2}\,p^{\alpha} \;\; \times \nonumber \\
 & &  \left[ \frac{p}{2y^{4}}  +\frac{1}{4}\,\bar{C}\,g(u)\,y^{-(3+\delta)}\,p^{\alpha-1} \;\;\times \right.\nonumber\\
& & \left. ( (4\alpha+\delta)\,y^{\delta/2}\, + 2\,\alpha\,\delta\,\bar{C}\,g(u)\,y\;p^{\alpha } ) \right].
\end{eqnarray}
It is easy to verify that both, $F_{y}\; > \; 0$ and $\Delta\; > \; 0$.

Once the location of the critical point, $u_{c}$, is known, we can solve $y_{c}$, $p_{c}$ and $\bar{C}$ from equations (\ref{Fcak}, \ref{Ucak}, \ref{Pcak}), obtaining:
\begin{equation}
y_{c}=\frac{1}{2}+\frac{\alpha}{\left(  1-\alpha\right)  }\frac{1}{q_{c}
}\left(  \frac{2}{u_{c}}+A\left(1+\Omega^{2}u_{c}\right)\right) \label{yc} \,,
\end{equation}
\begin{equation}
p_{c}=\frac{1}{2}q_{c}+\frac{\alpha}{1-\alpha}\left(  \frac{2}{u_{c}
}+A\left(1+\Omega^{2}u_{c}\right)\right) \label{pc} \,,
\end{equation}
\begin{equation}
\bar{C}=\frac{\,y_{c}^{\delta/2}}{\left(  1-\alpha\right)  g\left(
u_{c}\right)  p_{c}^{\alpha}}\left(  \frac{2}{u_{c}}+A\left(1+\Omega^{2}u_{c}\right)\right)
\label{Cc}\,,
\end{equation}
where $q_{c}$ is the positive solution of the quadratic equation
\begin{equation}
q^{2}+\frac{\delta}{1-\alpha}\left(  \frac{2}{u_{c}}+A\left(1+\Omega^{2}u_{c}\right)\right) q=\gamma\left(  u_{c}\right)
\label{eqq}\,,
\end{equation}
with
\begin{eqnarray}
\gamma\left( u_{c}\right)&=&-2 A \Omega^{2} + \frac{4}{u_{c}^{2}} \nonumber \\
& &+\frac{2h\left(u_{c}\right)  }{1-\alpha}\left(  \frac{2}{u_{c}}+A\left(1+\Omega^{2}u_{c}\right)\right).
\label{gammacakdelta}
\end{eqnarray}
Equations (\ref{yc}, \ref{pc} and \ref{Cc}) are generalizations
of the equations [49],[50] and[51] from Marlborough \& Zamir (1984), 
including now the effect of the line--force term $(n_{E}/W)^{\delta}$.

Since the eigenvalue $\bar{C}$ must be positive, we have:
\begin{equation}
\frac{2}{u_{c}} + A\left(1+\Omega^{2}u_{c}\right)\,>\,0 \,.
\end{equation} 
This last inequality imposes a restriction in the position (furthest from the stellar surface in the radial coordinate $r$) of the singular point, namely:
\begin{equation}
u_{c}\, < \,  u_{\max} \,\equiv\, -\frac{4}{A\left( 1+\sqrt{1-8\,\Omega^{2}/A}\right)} \,. \label{umax}
\end{equation}

\section{Topology of the rotating CAK wind \label{sec6}}
In this section we show the results of our topological analysis. Following Bjorkman (1995), we choose a typical $B2 \,V$ star with stellar parameters summarized in table \ref{tabla1}. We have adopted the line--force multiplier parameters $k$ and $\alpha$ from Abbott (1982) and the $\delta$ parameter is from LC, these values are summarized in table \ref{tabla2}.

\begin{table}
\caption[]{B2 V Stellar Parameters} \label{tabla1}
\begin{center}
\[
\begin{array}{cccccc}   %{p{0.5\linewidth}l}
\hline \hline
\noalign{\smallskip}
R/R_{\sun} \; & M/M_{\sun} \; & L/L_{\sun}  \; & T_{eff}/K\; &\Gamma\;& A\;\\
\noalign{\smallskip}
\hline
\noalign{\smallskip}
4.5\;&9.0\;&3553.\;&21000.\;&9.27\,10^{-3}\;&1413.\;\\
\noalign{\smallskip}
\hline
\end{array}
\]
\end{center}
\end{table}

\begin{table}
\caption{Line--force Parameters} \label{tabla2}
\begin{center}
\[
\begin{array}{ccc}   %{p{0.5\linewidth}l}
\hline \hline
\noalign{\smallskip}
k\;& \alpha \; & \delta \; \\%
\noalign{\smallskip}
\hline
\noalign{\smallskip}
0.212\; & 0.56\; & 0.02\; \\
\noalign{\smallskip}
\hline
\end{array}
\]
\end{center}
\end{table}

\subsection{The frozen--in ionization ($\delta=0$)}

The factor $(n_{E}/W)^\delta$ in the line--force takes into account the changes in the 
ionization of the  wind. As a first step to understand the topology of the rotating--CAK model
we set $\delta=0$. 
\subsubsection{The critical point interval}
This case is simpler because we can obtain analytically from equations (\ref{yc},\ref{pc},\ref{Cc}) 
the variable $y$ as function of $u$ and $p$, i.e.:
\begin{equation}
y=\frac{p\,}{2\,\left(  2/u+A(1+\Omega^{2}\,u)\,+p\,-C\,p^{\alpha}\,\right)  }\,.
\end{equation}
Bjorkman (1995) obtained the same result (see his equation [13]) for the
non--rotational case, $\Omega=0$. Moreover, $q_{c}$ can be expressed by:
\begin{equation}
q_{c}^2=\gamma(u_c)= 4 u_{c}^{-2} - 2 A \, \Omega^{2}  \label{gamacak}\,.
\end{equation}
As we mentioned in previous sections, $q_c$ must be positive. This gives an extra
restriction for the location of the singular point (nearest to the stellar
surface in the radial coordinate $r$). The value of $u_{min}$ is defined when 
$\gamma(u_{min})=0$ and is given by:
\begin{equation}
u_{\min} =\max\left\{  -1,-\frac{\sqrt{2/A}}{\Omega}\right\} \,.
\end{equation}
\begin{figure}[htbp] 
\begin{center}
\includegraphics[width=0.45\textwidth]{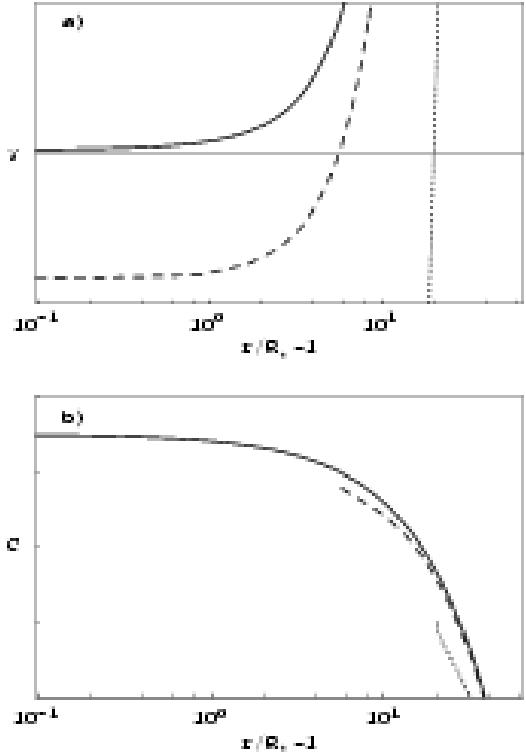}
\end{center}
\caption{a) The function $\gamma$ versus $r/R_{\ast}-1$ (eq. \ref{gamacak}) for 
different values of the rotational parameter $\Omega$. $\Omega=0$, 
continuous--line; $\Omega=0.25$, dashed--line; $\Omega=0.8$, dotted--line. The horizontal line is for $\gamma=0$.\newline
b) The eigenvalue from eq.~(\ref{Cc}) as a function of $r/R_{\ast}-1$ for the same values of $\Omega$ as a). 
Note that $d\bar{C}/du$ is always negative in the interval $(r_{min},r_{max})$. \label{gama_C_delta00}}
\end{figure} 
Therefore, $u_{c}$ is restricted to the interval $(u_{min}, u_{max})$. A similar result which 
restricts the location of the critical point to an interval has been found by 
Marlborough \& Zamir~(1984) see their eq. [65]).\\
Figure \ref{gama_C_delta00}a  shows the behavior of $\gamma$ (eq. \ref{gamacak}) versus
$r/R_{\ast}-1$  for different  values of the rotational parameter $\Omega$. For the non--rotational case 
$\Omega=0$, the function $\gamma$ is positive for almost the whole integration domain, therefore the singular point can be placed anywhere. Thus, is the lower boundary condition which fixes the position of the singular point. For the rotational case, the second term in the RHS of equation (\ref{gamacak}) is the dominant term for almost any value of $\Omega$. Therefore the larger is $\Omega$ the larger is the value of $r_{min}$ ($\equiv -1/u_{min}$) as the different curves in figure \ref{gama_C_delta00}a show. \newline

On the other hand, the value of the term $8 \Omega^2/A$ in eq.~(\ref{umax}) is almost negligible 
and consequently the value of $r_{max}$ ($\equiv -1/u_{max}$) is almost constant at $r=A R_{\ast}/ 2$. 
Table \ref{tabla3} shows the interval $(r_{min},r_{max})$ for different values of the parameter $\Omega$.
It is clear from this table that from a very low value of $\Omega=\sqrt{2/A}$ ($=0.038$ for our test star) the location of $r_{min}\,>\, R_{\ast}$ and the critical point, $r_{crit}$, is strongly shifted downstream in the wind. Once $r_{crit}$ (or $u_{c}$) is fixed, the value of $\gamma(r_{crit})$ is inserted in eq.~(\ref{gamacak}) and $q_{c}$ and $\bar{C}$ are obtained. Figure \ref{gama_C_delta00}b shows $\bar{C}$ from eq.~(\ref{Cc}) against $r/R_{\ast}-1$ for the same values of $\Omega$ of figure \ref{gama_C_delta00}a (same type of lines too). The value of the derivative, $d\bar{C}/du$ is always negative indicating that the critical point is an X--type.\\

\begin{table}
\caption{Analytical approximation and numerical results for the rotational--CAK model with $\delta=0$.
Note: the mass loss rate is given in units of  $10^{-9}\, M_{\sun} / year$ and the terminal velocity is in $km/sec$  \label{tabla3} }
\begin{center}
\[
\begin{array}{cccccccc}   %{p{0.5\linewidth}l}
\hline \hline
\noalign{\smallskip}
\Omega\; & r_{min}/R_{\ast} \;& r_{max}/R_{\ast} \;&r_{crit}/R_{\ast} \;& \bar{C}_{max} \;&  \bar{C} \;& \dot{M} \;&v_{\infty}\\%
\noalign{\smallskip}
\hline
\noalign{\smallskip}
0 \;& 1 \;& 706.50 \;& 1.559 \;& 48.248 \;& 48.223 \;& 1.814 \;& 1008.2\\
\hline
0.2\; &  5.316\; & 706.46 \;& 5.370  \;& 47.972 \;& 47.961\;& 1.831 \;& 920.1\\
0.4\; & 10.632\; & 706.34 \;& 10.669 \;& 47.648 \;& 47.634\;& 1.854 \;& 800.3\\
0.6\; & 15.948\; & 706.14 \;& 15.992 \;& 47.322 \;& 47.303\;& 1.877 \;& 683.3\\
0.8\; & 21.264\; & 705.86 \;& 21.334 \;& 46.992 \;& 46.964\;& 1.902 \;& 574.5\\
\noalign{\smallskip}
\hline
\end{array}
\]
\end{center}
\end{table}

\subsubsection{Linearization of the B--Matrix: Eigenvalues and Eigenvectors}
The matrix of linearization $B$ (eq.
\ref{mtxB}) is given by:
\begin{equation}
\left. B\,\right|  _{\left(  u_{c},y_{c},p_{c}\right)
 }\,=\,\frac{2}{u_{c}^{2}p_{c}}
\left(
\begin{array}[c]{cc}
\theta^{2} & \frac{\alpha\left(  1-\alpha\right)  u_{c}^{2}\eta}{2\left(
\theta+\alpha\eta\right)  }\\
\frac{2\left(  1+2\theta^{3}\right)  \left(  \theta+\alpha\eta\right)
}{u_{c}^{2}} & -2\theta^{2}
\end{array}
\right)
\end{equation}
where $\theta$ and $\eta$ are given by:
\begin{equation}
\theta=\sqrt{1-\Omega^{2}\,A\,u_{c}^{2}/2} \,,
\end{equation}
and 
\begin{equation}
\eta=-\left(
2+A\,u_{c}\,(1+\Omega^{2}\,u_{c})\right)  /\left(  1-\alpha\right)  \,.
\end{equation}
The eigenvalues of the $B$--matrix are:
\begin{equation}
\mu_{\pm}=\frac{1}{u_{c}^{2}p_{c}}\left(  -\theta^{2}\pm\sqrt{9\theta
^{4}+4\alpha\left(  1-\alpha\right)  \left(  1+2\theta^{3}\right)  \eta
}\right)\,.
\end{equation}
It is easy to check that the product of the eigenvalues
$\mu_{+}\times \mu_{-}$ is negative, i.e., the topology of the singular point
is saddle or X--type as the GH theorem establishes. We can write the associated eigenvectors as
$\left(  1,\nu_{\pm}\right)$, where 
\begin{equation}
\nu_{\pm}=\frac{\left(  -3\theta^{2}\pm\sqrt{9\theta^{4}+4\alpha\left(
1-\alpha\right)  \left(  1+2\theta^{3}\right)  \eta}\right)  \allowbreak
\left(  \theta+\alpha\eta\right)  }{\alpha\left(  1-\alpha\right)  u_{c}%
^{2}\eta}\, ,
\end{equation}
where $\nu_{+}$ ($\nu_{-}$) corresponds to the unstable (stable) manifold. 
The maximum value of $\bar{C}$, that accounts for the {\it{minimum}} mass loss rate,
occurs when $u_{c}=u_{\min}$, eq. (\ref{Cc}) becomes:
\begin{equation}
\begin{array} [l]{lll}
\bar{C}_{\max} =&\frac{A\,(1-\Omega^{2})-2}{\left(
 1-\alpha\right)  \,}\left(
\sqrt{1- \frac{\Omega^{2}\,A}{2}} \right. & \\
 & \left. +\frac{\,\alpha}{1-\alpha}\left(
A\,(1-\Omega^{2}\right)-2
 \right)  ^{-\alpha}& ,\; if \;\Omega\,\leq\,\sqrt{2/A}\\
\bar{C}_{\max} =& \frac{1}{\alpha^{\alpha}}\left(  \frac{A\,-2\,\sqrt{2 A}\,\Omega}
{1-\alpha}\right)  ^{1-\alpha} & , \;if\; \Omega \,> \, \sqrt{2/A}
\end{array}
\end{equation}

\subsubsection{Phase Diagram}
In order to obtain a numerical solution for the wind, we need an approximation for the location of 
the critical point. We use the value of $u_{c}=u_{min}$ as a first guess of the critical point 
and use this value of $u_{c}$ to calculate $\bar{C}_{min}$ as our first guest for the eigenvalue 
$\bar{C}$. Table \ref{tabla3} shows the values of $r_{min}$, $r_{crit}$ and $\bar{C}_{min}$, $\bar{C}$ 
confirming that this is a very good approximation. 
Now using standard numerical algorithms we integrate our system: $U=0$, $Y=0$ and $P=0$ 
(eqs. \ref{sist1}, \ref{sist2} and \ref{sist3}, respectively) from the singular point up and downstream to obtain the numerical solution.

\begin{figure}[htbp] 
\begin{center}
  \includegraphics[width=0.45\textwidth]{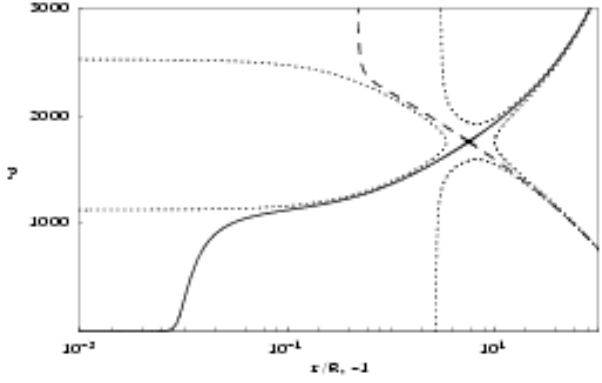} 
\end{center}
\caption{The topology of the freeze in ionization rotating--CAK model ($\delta=0$), $p$ versus $(r/R_{\ast}-1)$ for $\Omega=0.25$. The unique curve that starts at the stellar surface and reaches infinity is the CAK original solution (continuous--line). \label{phase_diag_delta00}}
\end{figure}

As Bjorkman~(1995) pointed out, it is insightfull to  study the solution topology in a $p$ versus 
$(r/R_{\ast}-1)$ diagram. Figure \ref{phase_diag_delta00} show this phase diagram.
If we start to integrate at the singular point, we cannot leave this point because all the equations, 
$U=0$, $Y=0$ and $P=0$ are simultaneously satisfied at this point. 
Therefore we have to move slightly up and downstream along the direction of the unstable manifold. After this, we can integrate obtaining the different solutions showed in figure \ref{phase_diag_delta00}. 
From this figure, it is clear that the only solution that reaches the stellar surface 
($\tau \rightarrow +\infty$, in the direction of $-(1,\nu_{+})$) and also reaches infinite ($\tau \rightarrow +\infty$, in the direction of $(1,\nu_{+})$) is the original CAK solution (continuous--line). The results from table \ref{tabla3} for the non--rotational case are the same one obtained by Bjorkman~(1995).

\begin{figure}[htbp] 
\begin{center}
  \includegraphics[width=0.45\textwidth]{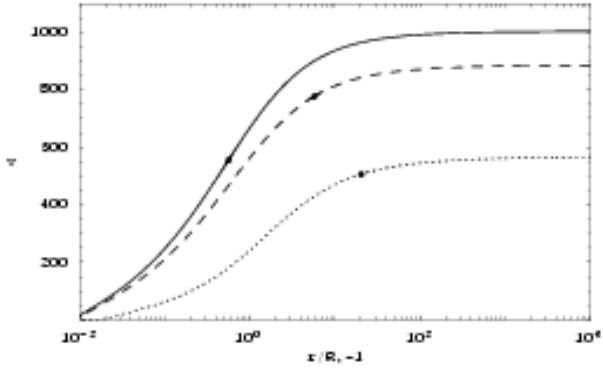} 
\end{center}
\caption{The velocity profile $v(r)$ (in $km/sec$) as function of  $r/R_{\ast}-1$ 
for the non rotating case (continuous--line) and for the rotational cases: $\Omega=0.25$ dotted--line; $\Omega=0.8$ dashed--line. The location of the singular points is indicated by a larger--dot \label{vel_delta00}}
\end{figure}
Figure \ref{vel_delta00} shows the velocity profile, $v$ (in $km/sec$) versus $r/R_{\ast}-1$  for our 
B2 \,V test star. We have chosen this value of $\Omega=0.25$ from the study of Abt et al.~(2002), 
that concluded that B--stars rotate at a $25\%$ of their break--up speed. The values $\Omega=0.8$ 
accounts for a fast rotator, e.g., a typical Be--Star (Chauville et al. 2001).  We see from this figure that neglects the rotational speed always overpredicts the value of the terminal velocity.\newline

We conclude that the rotational speed shifts the location of the critical point downstream and reduces the terminal velocity, but has almost no influence on the value of the eigenvalue (mass loss rate). 
Furthermore, we can see from our approximate and numerical results summarized in table \ref{tabla3} that the CAK wind {\it{do not}} have the {\it{smallest}} possible eigenvalue, $\bar{C}_{min}$, or the maximum mass--los rate as Feldmeier et al.~(2002) and  Owocki \& ud--Doula~(2004) concluded for a non--rotating CAK model with zero sound speed. Contrary to expectation, the rotating--CAK wind critical solution corresponds to an almost {\it{minimum}} mass--loss rate ({\it{maximum}} eigenvalue).

\subsection{The rotating CAK model ($\delta\neq 0$)}
Abbott~(1982) studied the indirect influence of the density in the line--force through the 
dependence of the ionization balance in the electron density. He found that this dependency 
modifies the force--multiplier and therefore the line--force by a factor $(n_{E}/W)^\delta$, 
where $n_{E}$ is the electron density (in units of $10^{-11}$ $gr\,/ cm^{3}$) and $W$ is the 
dilution factor.\\
Although this is a weak influence, because $\delta$ ranges between $0.0$ and $0.2$, it is 
important to study how its inclusion in the momentum equation modifies the topology of the 
rotating CAK model.\\

As we pointed out  in section \ref{sec5}, the existence of $q_{c}\,>\,0$ implies 
$\gamma\left(u_{c}\right)\,>\,0$. 
\begin{figure}[htbp] 
\begin{center}
  \includegraphics[width=0.45\textwidth]{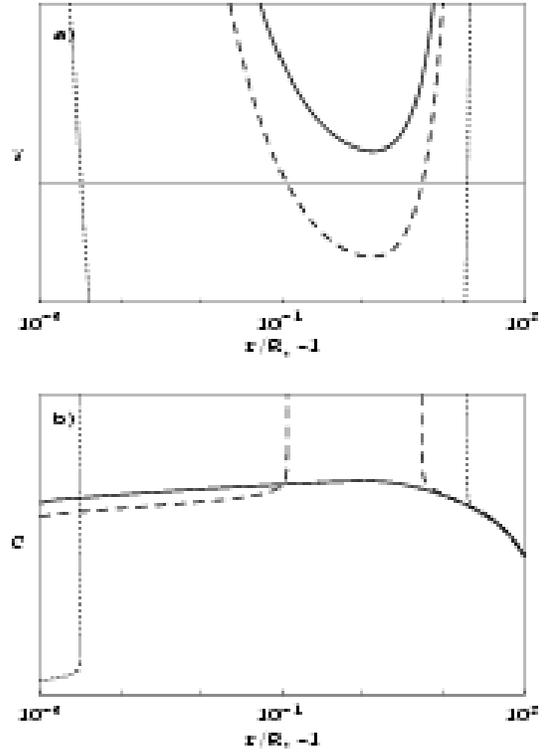} 
\end{center}
\caption{Same as figure \ref{gama_C_delta00} a) and b) but for $\delta=0.02$. See text for discussion. 
\label{gama_C_delta02}}
\end{figure}
The behavior of $\gamma$ (eq. \ref{gammacakdelta}) versus $r/R_{\ast}-1$ is shown in figure 
\ref{gama_C_delta02}a for the same values of $\Omega$ of figure \ref{gama_C_delta00}. We can 
interpret from this figure, that $\gamma$ in $(R_{\ast},r_{max})$ is a decreasing function in the 
neighborhood of $r \rightarrow  R_{\ast}^{+}$ and an increasing function in the neighborhood 
of $r \rightarrow r_{max}$. Furthermore, the function $\gamma$ posses only one minimum in the 
integration domain located at $r \equiv r_{min}$. The value of $\gamma(r_{min})$ decreases as $\Omega$ 
increases. For small values of $\Omega$, $\gamma(r_{min})\,>\,0$ and the location of the singular 
point $r_{crit}$ can be anywhere in the integration domain $(R_{\ast},r_{max})$. For larger values 
of $\Omega$, $\gamma(r_{min})\,<\,0$ and $\gamma(r)=0$ has two roots, at $r \equiv r_{\pm}$ where 
$r_{-}\,< \,r_{+}$ (or $u_{-}\,< \,u_{+}$). Thus, $r_{crit}$ can be located now in 
two different intervals, i.e., $r_{crit} \in (R_{\ast},r_{-})$ or $r_{crit} \in (r_{+},r_{max})$. 
For the particular case where $\gamma(r_{min})\,=\,0$, the value of the rotational speed parameter is 
$\Omega\,\equiv\,\Omega_{bif}$, where the subscript {\it{'bif'}} accounts for the bifurcation in 
the wind solution topology.\newline

Figure \ref{gama_C_delta02}b shows $\bar{C}$ (eq.~\ref{Cc}) against $r/R_{\ast}-1$ for different 
values of $\Omega$. When $\gamma(r)$ is positive, $\bar{C}$ exhibits now a different behavior 
compared with the $\delta=0$ case. As long as $\Omega < \Omega_{bif}$, $d\bar{C}/du$ is positive 
in the interval $(R_{\ast},r_{min})$, i.e., any singular point in this interval is an {\it{attractor}}. Furthermore, $d\bar{C}/du$ reaches its maximum, when $\gamma$ is minimum, i.e., in the neighborhood of
$r=r_{min}$, from this point up to $r_{max}$,  $d\bar{C}/du \,< \,0$ and any singular point in 
$(r_{min}, r_{max})$ is {\it{X--type}}. When $\Omega > \Omega_{bif}$, the minimum of $\gamma(r)$ is 
negative and the intervals are reduced to: $(R_{\ast},r_{-})$ for the attractor type singular point 
and $(r_{+}, r_{max})$ for the X--type singular point.\newline

The behavior of $y_{c}$, $p_{c}$ and $\bar{C}$ (eqs. \ref{yc}, \ref{pc} and \ref{Cc}) in the neighborhood of 
$r\,\rightarrow r_{\pm}$, in the inverse radial coordinate $u$, is as follows:
\begin{eqnarray}% [l]{ll}
y_{c} &  \sim        &\left|  u-u_{\pm}\right|  ^{-1}\,\rightarrow \, \infty \, ,\\
p_{c} &  \rightarrow &\frac{\alpha}{1-\alpha}\left(  \frac{2}{u_{\pm}} +A (1+\Omega^{2}u_{\pm})\right)  \, ,\\
\bar{C} &  \sim      &\left|  u-u_{\pm}\right|  ^{-\delta/2}\,\rightarrow \,+\infty \label{Crpm}
\end{eqnarray}
This divergent behavior of $\bar{C}$ in the neighborhood of $r_{\pm}$, is shown in the curves for 
$\Omega \ne 0$ in figure \ref{gama_C_delta02}. The almost constant value of $\bar{C}$ 
explains why a slightly change in the eigenvalue cause an enormous change in the location in the 
singular point. 

\begin{figure}[htbp] 
\begin{center}
  \includegraphics[width=0.45\textwidth]{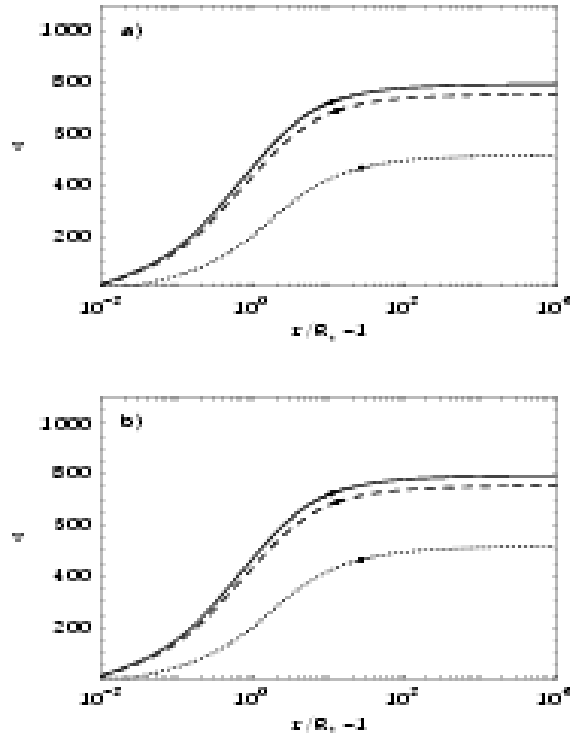}
\end{center}
\caption{The velocity profile $v(r)$ (in $km/sec$) as function of  $r/R_{\ast}-1$ 
for the non rotating case (continuous--line) and for the rotational cases: $\Omega=0.25$ dotted--line; $\Omega=0.8$ dashed--line. \label{vel_delta002-01}}
\end{figure}

Tables \ref{tabla4} and \ref{tabla5} summarise the numerical calculation for our test star with $\delta=0.02$  and $\delta=0.1$ respectively. The data of the $\dot{M}$ column (see also the $\bar{C}$ column) show that the effect of the rotation on $\dot{M}$ ($\bar{C}$) is almost negligible.
% for $\delta=0.02$ and small for $\delta=0.2$. 
Figure \ref{vel_delta002-01} shows the velocity profile for three different values of $\Omega$ ($0.0;0.25;0.8$), panel a) for $\delta=0.02$ and panel b) for $\delta=0.1$. 
A large dot shows the respective positions of the critical points. The effect of shifting the position of the critical point is stronger for low rotational speeds and decreases when $\Omega$ increases as a comparison between figures \ref{vel_delta00} and \ref{vel_delta002-01} clearly shows.
The terminal velocity, is a decreasing function of the rotational speed and has almost the same behavior as in the $\delta=0$ \,case. But for high rotational speeds, the influence of $\delta$ in $v_{\infty}$ is 
negligible/small as a comparison between table \ref{tabla3} and table \ref{tabla4}/\ref{tabla5} shows.\newline

We conclude from that the factor $(n_{E}/W)^\delta$ strongly shifts outwards the location of the critical 
point and produces a bifurcation in the solution topology.

\begin{table}
\caption{Numerical results for the rotation--CAK model with $\delta=0.02$.
Note: the mass loss rate is given in units of  $10^{-9}\, M_{\sun} / year$ and 
the terminal velocity is in $km/sec$ \label{tabla4} }
\begin{center}
\[
\begin{array}{ccccc}   %{p{0.5\linewidth}l}
\hline \hline
\noalign{\smallskip}
\Omega\; & r_{crit}/R_{\ast} \;& \bar{C} \;& \dot{M} \;&v_{\infty} \\
\noalign{\smallskip}
\hline
\noalign{\smallskip}
0 \;& 4.712 \;& 50.949 \;& 1.644 \;& 929.6 \\
\hline
0.2\;& 7.196  \;& 50.726\;& 1.657 \;& 879.3  \\
0.4\;& 11.926 \;& 50.291\;& 1.683 \;& 779.0 \\
0.6\;& 17.145 \;& 49.790\;& 1.713 \;& 669.9 \\
0.8\;& 22.576 \;& 49.247\;& 1.747 \;& 564.6 \\
\noalign{\smallskip}
\hline
\end{array}
\]
\end{center}
\end{table}

\begin{table}
\caption{Same as Table (\ref{tabla4}) but for $\delta=0.1$. \label{tabla5} }
\begin{center}
\[
\begin{array}{ccccc}   %{p{0.5\linewidth}l}
\hline \hline
\noalign{\smallskip}
\Omega\; & r_{crit}/R_{\ast} \;& \bar{C} \;& \dot{M} \;&v_{\infty} \\
\noalign{\smallskip}
\hline
\noalign{\smallskip}
0 \;& 11.708 \;& 63.099\;& 1.122 \;& 801.1 \\
\hline
0.2\;& 13.168 \;& 62.734\;& 1.133 \;& 775.1 \\
0.4\;& 16.873 \;& 61.876\;& 1.162 \;& 708.5 \\
0.6\;& 21.853 \;& 60.619\;& 1.205 \;& 621.1 \\
0.8\;& 27.683 \;& 59.107\;& 1.261 \;& 528.5 \\
\noalign{\smallskip}
\hline
\end{array}
\]
\end{center}
\end{table}

\subsubsection{Bifurcation rotational speed \label{sec-CRBI}}
In order to have an analytical approximation for the value of $\Omega_{bif}$, 
we can approximate for the function $h(u)$, eq. (\ref{eqh}):
\begin{equation}
h(u) \simeq -\frac{\delta }{2 }u \,,
\end{equation}
then the minimum of $\gamma(u)$ is achieved at:
\begin{equation}
u_{bif} \simeq -2\left( \frac{1-\alpha }{\delta A}\right) ^{1/3} \,,
\end{equation}
and the minimum value of $\gamma(u_{bif})$ is 
\begin{eqnarray}
\gamma(u_{bif}) &\simeq &-2\,\left( 1-\alpha \right)
\,\Omega^{2} \,A \, \left( 1+2\left( 1-\alpha \right) ^{-1/3}\delta
^{1/3}A^{-2/3}\right) \nonumber \\
& & +3\,\,\left( 1-\alpha \right) ^{1/3}\delta ^{2/3}A^{2/3}-2\,\delta \,
\end{eqnarray}
From $\gamma(u_{bif}) = 0$, we can obtain the bifurcation value of 
$\Omega$ given by 
\begin{equation}
\Omega_{bif} \simeq \sqrt{\frac{3\,\,\left( 1-\alpha \right) ^{1/3}\delta
^{2/3}A^{2/3}-2\,\delta \,}{2\,\left( 1-\alpha \right) \, A \,\left( 1+2\left(
1-\alpha \right) ^{-1/3}\delta ^{1/3}A^{-2/3}\right) }} \label{omega_bif_app}
\end{equation}

\begin{figure}[htbp] 
\begin{center}
  \includegraphics[width=0.45\textwidth]{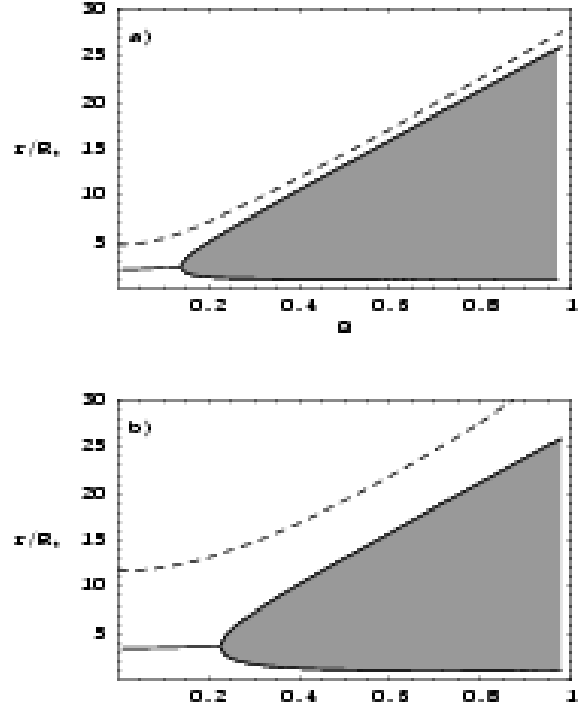}
\end{center}
\caption{The curves $r_{min}$ and $r_{\pm}$ are shown with continuous lines as function of $\Omega$. 
These curves represent the boundaries for the type of the singular point topology.
The numerical results for the location of the  singular point is also shown in dashed--line. 
Figure a) is for $\delta=0.02$ and b) for $\delta=0.1$. See text for details. \label{Rcrit_delta_omega}}
\end{figure} 

Figure  \ref{Rcrit_delta_omega} show the curves $r_{min}$ and $r_{\pm}$ as function of $\Omega$. 
The intersection point for all the curves (in continuous--line) is at $\Omega=\Omega_{bif}$. Critical 
points can not be located between the curves $r_{+}$ and $r_{-}$ (filled region). In addition, we show 
curves for the  location of the critical point from numerical calculations (dashed--lines), with the 
lower boundary condition, $\tau_{Phot}=2/3$. \newline

We can clearly see from this figure, that the position of the critical point is shifted outwards from 
the stellar surface and the greater is $\delta$ the further is the position of this critical point.
It can be inferred from this figure, that the location of the singular point remains almost constant as long as $\Omega \, \le \, \Omega_{bif}$, but from values of $\Omega \, > \, \Omega_{bif}$ the position of $r_{crit}$ grows almost linear with $\Omega$. 

This behavior of the solution topology can be applied for the winds of Be--Stars. At polar latitudes, i.e., slow rotational speed, the wind behaves as the standard CAK wind, but as the latitude approaches to the equator, the rotational speed is larger than $\Omega_{bif}$ and the wind is slower and denser. This transition from polar to equatorial latitudes seems to have a similar behavior described by Cur\'{e}~(2004) for the more general rotating m--CAK wind. The study of the influence of this bifurcation in the winds of Be stars will be the scope of a forthcoming article.

\subsubsection{Phase Diagram with $\delta\neq 0$ . \label{secphddelta}}
\begin{figure}[htbp] 
\begin{center}
  \includegraphics[width=0.45\textwidth]{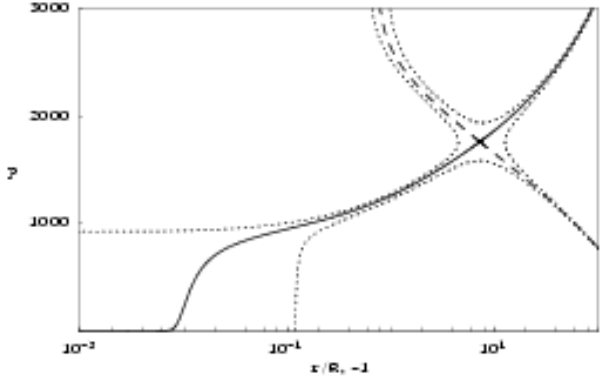} 
\end{center}
\caption{The topology of the rotating--CAK model, $p$ versus $r/R_{\ast}-1$ 
for $\Omega=0.25$ and $\delta=0.02$. The unique curve that starting at the stellar surface 
and reaches infinity is the CAK original solution (continuous--line). \label{phase_diag_delta02}}
\end{figure}

\begin{figure}[htbp] 
\begin{center}
  \includegraphics[width=0.45\textwidth]{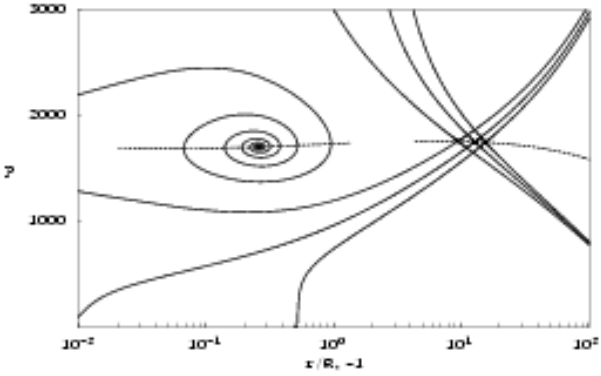} 
\end{center}
\caption{The topology of the rotating--CAK model, $p$ versus $(r/R_{\ast}-1)$ 
for $\Omega=0.25$. The different curves correspond to different values of the eigenvalues.
The left--most curve starts at the 'normal' critical point and is trapped by the attractor. See text for details.
 \label{phase_diag_attractor}}
\end{figure}

Figure \ref{phase_diag_delta02} shows the phase diagram $p$ versus $r/R_{\ast}-1$ for $\Omega=0.25$ and $\delta=0.02$ for our test star. The solution topology seems to be similar to the $\delta=0$ case. Here the shallow and steep curves are in continuous line and dashed line respectively. The shallow solution is the CAK solution while the steep solution correspond to accretion flows or for radiation driven disk winds (Feldmeier et al.~2002). As we mentioned in previous section, we move slightly from the singular point in the unstable manifold in the directions $\pm(1,\nu_{+})$ and then integrate outwards and inwards, obtaining the solution topology of figure \ref{phase_diag_delta02}. 
Figure \ref{phase_diag_attractor} shows both shallow and steep solutions, but here we have started from the singular point but with different values of the eigenvalue $\bar{C}$. The almost horizontal dotted curves correspond to $p_{c}$ from eq. (\ref{pc}) and is not a continuous curve because of the bifurcation in the solution topology. The left--most shallow curve that start at a X--type singular point do not reach stellar surface, but is trapped by the {\it{attractor}} singular point. This result implies that there is a shorter window of locations of singular points, that fixes the eigenvalues $\bar{C}$ and can reach infinity passing through the X--type critical point.
This feature is not present when $\delta=0$.

\section{Conclusions \label{sec-CONC}}
In this article we have examined the topology of the rotating--CAK wind non--linear differential equation.
After the introduction of an additional physically meaningless independent variable ($\tau$) we transform the momentum equation to a set of equations where all the derivatives are explicitly given. This formalism permitted us to linearise the equations in the neighborhood of the critical points. This linearization let us to define a condition for the derivative of the eigenvalue that defines the topology of the critical point, i.e, X--type or attractor.

We have applied our results to the case of a point star (CAK) for a frozen in ionization rotating wind, recovering and generalizing the results of previous studies (Bjorkman~1995 and Marlborough \& Zamir~1984). The most significant result (with $\delta = 0$) is that the wind {\it{ does not}} assume the maximum mass-loss rate but almost the {\it{minimum}}.

For the more general case, where changes in the wind ionization are taken into account, our analysis shows the existence of a bifurcation in the solution topology, where two critical points exist.
The first critical point (closer to the star's surface) is an attractor while the second is the standard CAK critical point. Besides the known fact that the rotational speed shifts the location of the critical point outwards in the wind, the inclusion of the term $(n_{E}/W)^{\delta}$ produces the same effect, reinforcing this displacement.

The bifurcation topology seems to explain the results from Cur\'{e}~(2004) that there exist two regions in the wind of a fast rotating hot star, one where the wind is the one from the standard solution (fast wind) and the other with a new solution that is slower and denser. This result shows us the necessity to perform a topological analysis of the rotating m--CAK wind. This study is currently underway.

\appendix
\section{Linearization in the neighborhood of a singular point}
Considering  the system of differential equations given by:%
\begin{eqnarray}
\dot{x}= dx/dt  & =X\left(  x,y\right)  ,\nonumber\\ 
\dot{y}= dy/dt  & =Y\left(  x,y\right)  ,%
\end{eqnarray}
where $x$ and $y$ are the dependent variables and $t$ is the independent variable.
The point $\left(  x_{c},y_{c}\right)$ is singular (critical) if and only if
verifies that:
\begin{eqnarray*}
X\left(  x_{c},y_{c}\right)&  =0 \nonumber\\
Y\left(  x_{c},y_{c}\right)&  =0 \nonumber
\end{eqnarray*}
In order to understand the topological behavior of the solutions close to
a singular point, we expand this system using Taylor series at $\left(
x_{c},y_{c}\right)$, obtaining:
\begin{eqnarray*}
\dot{x}  & =X|_{\left(  x_{c},y_{c}\right)  }+X_{x}|_{\left(  x_{c}%
,y_{c}\right)  }\left(  x-x_{c}\right)  +X_{y}|_{\left(  x_{c},y_{c}\right)
}\left(  y-y_{c}\right)  +o\left(x,y\right)  \\
\dot{y}  & =Y|_{\left(  x_{c},y_{c}\right)  }+Y_{x}|_{\left(  x_{c}%
,y_{c}\right)  }\left(  x-x_{c}\right)  +Y_{y}|_{\left(  x_{c},y_{c}\right)
}\left(  y-y_{c}\right)  +o\left(x,y\right)
\end{eqnarray*}
Neglecting superior order terms and using that $\left(  x_{c},y_{c}\right)  $
is a singular point, we have, in matrix form:
\begin{equation}
\left(
\begin{array}
[c]{c}%
\dot{x}\\
\dot{y}%
\end{array}
\right)  =B  \left(
\begin{array}
[c]{c}%
x-x_{c}\\
y-y_{c}%
\end{array}
\right)
\end{equation}
where $B$ is the Jacobian matrix at $\left(  x_{c},y_{c}\right)  $, i.e.,
\begin{equation}
B=\left.  \left(
\begin{array}
[c]{cc}%
X_{x} & X_{y}\\
Y_{x} & Y_{y}%
\end{array}
\right)  \right|  _{\left(  x_{c},y_{c}\right)  }%
\end{equation}
For a 3-dimensional system case:
\begin{eqnarray}
\dot{x}  & =X\left(  x,y,z\right)  \nonumber \, ,\\
\dot{y}  & =Y\left(  x,y,z\right)  \nonumber\, , \\
\dot{z}  & =Z\left(  x,y,z\right)  \label{3-d}\, ,
\end{eqnarray}
with a constraint $\phi\left(  x,y,z\right)$ satisfying:
\begin{equation}
\phi_{x}X+\phi_{y}Y+\phi_{z}Z = 0 \, ,
\end{equation}
the surface defined by:
\begin{equation}
S=\left\{  \left(  x,y,z\right)  :\phi\left(  x,y,z\right)  =0\right\} , \label{3-d-rest}
\end{equation}
is invariant for the differential equation system evolution. 

From the equation (\ref{3-d-rest}), we can solve for $z$ (using the implicit function theorem) obtaining $z=z\left(  x,y\right)$ and reduce the system (\ref{3-d}) to:
\begin{eqnarray}
\dot{x}  & =X\left(  x,y,z\left(  x,y\right)  \right)  \, , \nonumber \\
\dot{y}  & =Y\left(  x,y,z\left(  x,y\right)  \right)  \, .\label{2-d}
\end{eqnarray}

If $\left(  x_{c},y_{c},z_{c}\right)  $ is a singular point of (\ref{2-d}), the Jacobian matrix $B$ is
given by:
\begin{equation}
B=\left.  \left(
\begin{array}
[c]{cc}%
X_{x}+z_{x}X_{z} & X_{y}+z_{y}X_{z}\\
Y_{x}+z_{x}Y_{z} & Y_{y}+z_{y}Y_{z}%
\end{array}
\right)  \right|  _{\left(  x_{c},y_{c}\right) } .%
\end{equation}

Since $z_{x}=-\phi_{x}/\phi_{z}$ and $z_{y}=-\phi_{y}/\phi_{z}$, we have
\begin{equation}
B=\frac{1}{\phi_{z}}\left.  \left(
\begin{array}
[c]{cc}%
\phi_{z}X_{x}-\phi_{x}X_{z} & \phi_{z}X_{y}-\phi_{y}X_{z}\\
\phi_{z}Y_{x}-\phi_{x}Y_{z} & \phi_{z}Y_{y}-\phi_{y}Y_{z}%
\end{array}
\right)  \right|  _{\left(  x_{c},y_{c}\right)  } .%
\end{equation}
For $\phi=F$, $X=U$ and $Y=P$, we obtain equation (\ref{mtxB}).

\begin{acknowledgements}
This work has been possible thanks to the research cooperation agreement UBA/UV and DIUV project 15/2003. MC wants to thank the hospitality of the colleges from the mathematics department from the UBA during his stay in Buenos Aires.
\end{acknowledgements}

\end{document}